\documentstyle[12pt, epsf]{article}

\newcommand{\be}{\begin{equation}}
\newcommand{\ee}{\end{equation}}
\newcommand{\beqs}{\begin{eqnarray}}
\newcommand{\eeqs}{\end{eqnarray}}

\def\({\left(}
\def\){\right)}

\def\d{\delta}

\def\a{\alpha}

\def\th{\theta}

\def\ni{\noindent}

\def\f{\frac}

\def\N{${\cal N}$}
\def\p{\phi}
\def\pb{\bar{\phi}}
\def\sym{SYM$_4\;\;\;$}
\def\nn{\nonumber}
\def\ds{D^2}
\def\dbs{{\bar{D}}^2}
\def\J{$J$}
\def\Jb{$\bar{J}$}
\def\ni{\noindent}

\begin{document}

\begin{titlepage}

\begin{flushright}
\begin{tabular}{l} ITP-SB-99-5 \\UMDEPP-99-97\\ hep-th/9903094 \\ March, 1999
\end{tabular}
\end{flushright}

\vspace{8mm}
\begin{center} {\Large \bf Non-renormalization of two and three Point Correlators of \N=4 SYM in ${\cal N}$=1 Superspace }

\vspace{20mm}

F.~Gonzalez-Rey$^{\dagger}$ 
          \footnote{email: glezrey@bouchet.physics.umd.edu},
B.~Kulik$^{\ddagger}$ \footnote{email: bkulik@insti.physics.sunysb.edu},
I.Y.~Park$^{\ddagger}$  \footnote{email: ipark@insti.physics.sunysb.edu},

\vspace{10mm} 
$^{\dagger}$Elementary Particle Group\\
Physics Department\\
University of Maryland\\
College Park, MD 20742\\
\vspace{2mm} 
$^{\ddagger}$Institute for Theoretical Physics \\ 
State University of New York	\\ 
Stony Brook, N. Y. 11794-3840 \\

\vspace{20mm}

\begin{abstract}
Certain two and three point functions of gauge invariant primary operators of \N=4 SYM are computed in \N=1 superspace keeping all the $\th$-components. This allows one to read off many component descendent correlators. Our results show the only possible $g^2_{YM}$ corrections to the free field correlators are contact terms. Therefore they vanish for operators at separate points, verifying the known non-renormalization theorems. This also implies the results are consistent with \N=4 supersymmetry even though the Lagrangian we use has only \N=1 manifest supersymmetry. We repeat some of the calculations using supersymmetric Landau gauge and obtain, as expected, the same results as those of supersymmetric Feynman gauge.   
\end{abstract}

\end{center}

\vspace{35mm}

\end{titlepage}
\newpage
\setcounter{page}{1}
\pagestyle{plain}
\pagenumbering{arabic}
\renewcommand{\thefootnote}{\arabic{footnote}} \setcounter{footnote}{0}

\section{Introduction}      
Perturbative string theories are described by two dimensional world sheet conformal field theories: String theory is holographic\cite{TS} at the perturbative level. Maldacena's conjecture\cite{Mal} states that this is true even at the non-perturbative level. According to this conjecture, the physics of string/M theory on various near horizon geometries is encoded in lower dimensional conformal field theories. One of the  examples considered is D3 branes in type IIB superstring theory. By taking a limit where the field theory on the brane decouples from the bulk, it was argued that type IIB string/supergravity on AdS$_5\times$ S$_5$ is dual to D=4, \N=4 super Yang-Mills(SYM$_{\small{4}}$). Therefore, if the conjecture is correct in its strongest form, i.e., for any $N_c$ and $g^2_{YM}$, it could provide non-perturbative understanding of type IIB string theory through its connection to SYM$_4$. This connection is made more precise in \cite{GKPW}. There it was proposed that bulk configurations with some boundary conditions should be described by SYM$_4$ with appropriate source terms couple to the bulk fields evaluated on the boundary. There has been progress in understanding AdS dynamics from SYM$_4$\cite{BD}.\\       
\indent On the other hand, in a large $N_c$ with large fixed $\lambda=g^2_{YM}N_c$, supergravity can be used to describe strongly coupled SYM$_4$. Much evidence has been accumulated so far along this line\cite{many}. In \cite{S}, in particular, all three point functions of general gauge invariant 
chiral primary operators (CPO) of \N=4 SYM were considered keeping only free field 
contributions and compared with the supergravity calculations. By the equality
of those two results, the authors were led to conjecture that the three point 
correlators of all CPOs at large $N_c$ are independent 
of $\lambda$. To the order of $g^2_{YM}$, D'Hoker et al. \cite{DFS} checked the conjecture (even at finite $N_c$) by exploiting the known non-renormalization theorems \cite{Nonre1} for certain two and three point functions.\\ 
\indent In both directions of the studies of AdS/CFT correspondence mentioned above, it is important to have an efficient tool to compute the SYM correlators of SYM$_4$. In this letter, we present a \N=1 superspace computation of $g^2_{YM}$ corrections to certain two and three point functions of gauge invariant operators(Four point functions have been considered in \cite{GPS} keeping the lowest components and in \cite{West} using \N=2 harmonic superspace.). The full \N=1 computation does not require much additional work and provides more data, i.e., descendent operators, at one stroke. This work is a first step toward the calculation of more general correlators than the ones considered here and correlators at orders higher than $g^2_{YM}$. For the two and three point functions considered, we show the only possible leading corrections are contact terms and therefore vanish for the operators at separate points verifying the known non-renomalization theorems. This shows our results are consistent with the \N=4 supersymmetry \cite{DFS} even though the calculation has been done in \N=1 superspace. A harmonic superspace discussion appears in \cite{HW}.

In section 2, we compute at the order of $g^2_{YM}$ the amplitudes of products of the traceless operators  
$<O^{11}O_{11}>\equiv\left< \;tr\(\phi^1 \phi^1\) \;tr(\bar{\phi}_1\bar{\phi}_1)\;\right>$,
$<{O^1}_2 {O_1}^2>\equiv<tr(\p^1e^V\pb_2) tr(\pb_1e^V\p^2) >$, and discuss $R$-symmetry invariance of the results. We also compute $<O^{11}O^{22}O^{33}>\equiv <tr\(\p^1\p^1\)tr\(\p^2\p^2\)tr\(\p^3\p^3\)>$. In section 3, we discuss the calculations in supersymmetric Landau gauge and check the gauge choice independence. Implications of our results are discussed in the concluding section. Throughout this article, we follow the conventions of \cite{book}.         

\section{Calculations of Two and Three Point Correlators}

As in \cite{GPS}, we add the appropriate source terms to the \sym action,

\beqs
{\cal L} &=& {\cal L}_{\small{\mbox{cl}}} + {\cal L}_{J} \nonumber \\ 
         &=&\f{1}{g^2_{YM}} \;tr\(\;\int d^4xd^4\theta\; e^{-V}\pb_i e^V\p^i\;
                              +\;\int d^4xd^2\theta\;W^2 \right. \nn \\
         & &   \left. +\f{1}{3!}\int d^4xd^2\theta iC_{ijk}\p^i[\p^j,\p^k]
                               \;+\;\f{1}{3!}\int d^4xd^2\bar{\theta}
                                iC^{ijk}\pb_i[\pb_j,\pb_k]  \right.\nn\\
         & &   \left. +\int d^4xd^2\theta\; J_{ij}tr\p^i\p^j
                      +\int d^4xd^2\bar{\th}\;\bar{J}^{ij}tr\pb_i\pb_j
                      +\int d^4xd^4\th\;{J_i}^j tr\p^ie^V\pb_j  \; \), \nn\\
                                      \label{L}
\eeqs  

\ni where $\p$ is an ${\cal N}=1$ chiral superfield containing $(\p,\psi_{\alpha},F)$ and vector superfield $V$ contains $(A_{\a\dot{\a}},\lambda_\a,D')$ in the Wess-Zumino gauge. Whether $\p$ is a superfield or a component will be obvious from the context. The superspace correlators can be obtained by first computing the effective action of $J$'s, $S_{eff}(J)$, and then taking the functional derivatives with respect to $J$'s. One can also read off many component correlators from $S_{eff}(J)$ by taking functional derivatives with respect to component sources which, in the chiral case, are defined by

\beqs
J_{\p}   & \equiv &  J|  \nn\\
J_{\psi} & \equiv &  DJ| \nn\\
J_{F}    & \equiv &  \ds J|
\eeqs
\ni and similarly for complex conjugates. One can define the components of a non-chiral source too, but below we will consider only chiral cases to illustrate this point. 
 
The easiest example to compute is $<O^{11}O_{11}>=<tr\p^1\p^1\,tr\pb_1\pb_1>$. All the two-loop contributions are given in Fig.~\ref{fig:2all1}.

\begin{figure}[!ht]
\centerline{
        \begin{minipage}[b]{2.5cm}
                \epsfxsize=2.5cm
                \epsfbox{2all2c.epsi}
        \end{minipage}
        \begin{minipage}[b]{2.5cm}
                \epsfxsize=2.5cm
                \epsfbox{2all1a.epsi}
        \end{minipage}
        \begin{minipage}[b]{2.5cm}
                \epsfxsize=2.5cm
                \epsfbox{2all2a.epsi}
        \end{minipage}           
        \begin{minipage}[b]{2.5cm}
                \epsfxsize=2.5cm
                \epsfbox{2all1b.epsi}
        \end{minipage}
}
\caption{Graphs for $<O^{11}O_{11}>$}
\label{fig:2all1}
\end{figure}   

\ni In the Feynman gauge the only contribution comes from the graph on Fig.~\ref{fig:2all1}a. The self-energy insertions, Fig.~\ref{fig:2all1}b,d,
cancel each other. The tadpole graph, Fig.~\ref{fig:2all1}c~, vanishes in Feynman gauge. Writing down covariant derivatives explicitly on Fig.~\ref{fig:2all1}a~, and absorbing $D^2,{\bar{D}}^2$ to complete the superspace measures of the chiral vertices, we have

\newpage

\begin{figure}[!ht]
\centerline{ 
\epsfxsize=9cm
\epsfbox{11iit.epsi}
}
\caption{}
\label{fig:11iib}
\end{figure}

\ni where the equality is obtained by partially integrating the $\dbs,\ds$ in the bottom lines as indicated by arrows. First consider $\dbs$. There are three terms, but the only contribution comes from when $\dbs\;$ acts on $\bar{J}$: The other two graphs vanish because of the Grassman $\delta$-functions on $k$-line and $q$-line. A similar argument holds for $\ds$ and the result is

 \beqs        
S^{(2)}_{eff}
        &=& 2
                             \int{d^4pd^4\theta} 
                     J(-p,\theta)_{11}D^2\bar{D}^2
                          \bar{J}^{11}(p,\theta)\frac{A}{-p^2}            
\eeqs       
   
\ni where the superscript on $S$ indicates it is a two loop result and $A$ is a finite number defined by,

\beqs
       \f{A}{p^2}   &\equiv&    \int d^4qd^4k~
                     \f{1}{q^2}
                         \f{1}{k^2}
                           \f{1}{(p+k)^2}
                        \f{1}{(p-q)^2}
                         \f{1}{(q+k)^2}  \nn \\ 
               &=&  \f{1}{p^2} \int d^4xd^4y~
                     \f{1}{x^2}
                         \f{1}{y^2}
                           \f{1}{(n+x)^2}
                        \f{1}{(n-y)^2}
                         \f{1}{(x+y)^2} \nn  \label{A}
\eeqs
\ni where $n_\mu = p_\mu/|p|$ and the propagators are Euclidean ones. 
We omit the overal factor $g^2_{YM} N_c(N_c^2-1)$ because all corrections we compute have the same factor. From now on, we will often drop sub- and/or super- indices on $J$'s too. 
Since the sources $\bar{J}$ are anti-chiral, we can replace $\ds\dbs$ by $-p^2$:

$$
 S^{(2)}_{eff}(J)= 
2  A
\int{d^4pd^4\theta J(-p,\theta)\bar{J}(p,\theta)}
$$
  
\ni Upon Fourier transform, it becomes a local expression

\be
S^{(2)}_{eff}(J) = 
2A
\int{d^4xd^4\theta J(x,\theta)\bar{J}(x,\theta)} \label{jjb}
\ee

\ni Taking functional derivatives with respect to \J,\Jb, we obtain

\be
<\;tr\(\p^1(z_1)\phi^1(z_1)\)\;tr\(\pb_1(z_2)\pb_1(z_2)\)>\;=\; 2A\;\int d^8z
                   \d^8(z_1-z)\dbs\ds \d^8(z-z_2)  \label{O11O11b}
\ee

\ni where $z=(\th,x)$. From (\ref{jjb}) we can read off many component correlators. To do that, first consider the relevant source terms in components. In this case, the relevant source terms are chiral ones and antichiral ones. In a schematic notation, we have

\be
\int \d^4x d^2\th {\cal L}_{J}= \int d^4x \;
                 \left[   J_{F}\,\p\p+2J_{\psi}\p\psi
                      +J_{\p}(2\p F+\psi\psi) \right]\;+\;h.c.
\ee

\ni Therefore taking functional derivatives of path integrals with respect to two of the three component sources gives a corresponding two point function. For example, we obtain, in a schematic notation,

\beqs
\f{\d}{\d J_F}\f{\d}{\d \bar{J}_{\bar{F}}}S^{(2)}_{eff}
             &\sim&  <\p \p \;\pb \pb>\;  \sim  \;\d(x_1-x_2)\nn\\
\f{\d}{\d {J}_{{\psi}}}\f{\d}{\d\bar{J}_{\bar{\psi}}}S^{(2)}_{eff}
             &\sim&  <\p\psi\;\pb\bar{\psi}>\; 
                        \sim  \;\partial\d(x_1-x_2)\nn\\
\f{\d}{\d J_{\p}}\f{\d}{\d J_{\pb}}  S^{(2)}_{eff}
             &\sim& <(2F\p+\psi\psi)(2\bar{F}\pb+\bar{\psi}\bar{\psi})>\;
                        \sim \; \Box \d (x_1-x_2)  \label{comp}          
\eeqs

\ni where $\sim$ indicates that the equations are true up to unimportant numerical factors. The contact terms are consistent with the conformal symmetry. All the other correlators are zero. For example,

\beqs
\f{\d}{\d J_F}\f{\d}{\d \bar{J}_{\bar{\p}}}S^{(2)}_{eff}
             &\sim&  <\p \p \;(2\bar{F}\pb+\psi\psi)>\;  = 0\nn\\
\f{\d}{\d {J}_{{\psi}}}\f{\d}{\d\bar{J}_{\bar{F}}}S^{(2)}_{eff}
             &\sim&  <\p\psi\;\pb\pb>\;  =  \; 0  \nn\\                
\eeqs
etc. The correlators in (\ref{comp}) also  vanish for $x_1\neq x_2$, therefore consistent with the known non-renormalization theorems. Obviously this is true for all the other component correlators.

The next correlator, 
$<{O^1}_2{O_1}^2>$, has extra graphs in addition to those in Fig.~\ref{fig:2all1}. As before, all self-energy graphs and tadpoles do not contribute. The graphs we need to consider are

\begin{figure}[!ht]
\centerline{
        \begin{minipage}[b]{2.5cm}
                \epsfxsize=2.5cm
                \epsfbox{2all5a.epsi}
        \end{minipage}
        \begin{minipage}[b]{2.5cm}
                \epsfxsize=2.5cm
                \epsfbox{2all5b.epsi}
        \end{minipage}
        \begin{minipage}[b]{2.5cm}
                \epsfxsize=2.5cm
                \epsfbox{2all5c.epsi}
        \end{minipage}
         \begin{minipage}[b]{2.5cm}
                \epsfxsize=2.5cm
                \epsfbox{2all5d.epsi}
        \end{minipage}          
}
\caption{Graphs for $<O^1_2O_1^2>$}
\label{fig:2all5}
\end{figure}

\ni One must add all these four diagrams with appropriate combinatoric factors. After some algebra, one obtain the result


\begin{figure}[!ht]
\centerline{
\epsfxsize=9cm
\epsfbox{1212rt2.epsi}
}
\caption{}
\label{fig:1212r2}
\end{figure}
   
\ni which corresponds to

\beqs
S^{(2)}_{eff} 
        &=& 
                \int{d^4pd^4\theta J(-p,\theta)(-p^2)(\dbs\ds + 
                 \frac{1}{2}p_{\alpha \dot{\alpha}}
               \bar{D}^{\dot{\alpha}} D^{\alpha})J(p,\theta) 
                      } \nn\\
        & &     \int{
                \int{d^4kd^4q \frac{1}{q^2}
                    \frac{1}{k^2}
                    \frac{1}{(p+k)^2}
                    \frac{1}{(p-q)^2}
                    \frac{-1}{(q+k)^2}   }} \nn\\
        &=&        
                \int{d^4pd^4\theta 
                    J(-p,\theta)(-p^2)(\dbs\ds 
                       + \frac{1}{2}p_{\alpha \dot{\alpha}}
                       \bar{D}^{\dot{\alpha}} D^{\alpha} )J(p,\theta)
                       \frac{A}{(-p^2)} } \nn\\    
        &=&     A
                          \int{d^4pd^4\theta      
                      J(-p,\theta)(\dbs\ds 
                         + \frac{1}{2}p_{\alpha \dot{\alpha}}
                  \bar{D}^{\dot{\alpha}} D^{\alpha}  ) J(p,\theta) }
\eeqs
\ni where $A$ is the same as in eq(\ref{A}). Fourier transformation to $x$-space and functional differentiation give the following result:

\beqs
            S^{(2)}_{eff} & =&    A
                            \int{d^8z 
                           {J_1}^2(z)(\dbs\ds 
                           + \frac{i}{2}\partial_{\alpha \dot{\alpha}}
                            \bar{D}^{\dot{\alpha}} D^{\alpha}  )     
                             {J_2}^1(z) } \nn\\   
                          & =&    A
                            \int{d^8z 
                           {J_1}^2(z)\( \dbs\ds 
                           - \frac{1}{2}\Box  \)     
                             {J_2}^1(z) } 
                                      \label{j12j12}
\eeqs

\beqs 
\f{\d}{\d J(z_1)}\f{\d}{\d J(z_2)}  
             S^{(2)}_{eff} 
                     &=&   <tr\(\p^1(z_1)\pb_2(z_1)\)
                            tr\(\pb_1(z_2)\p^2(z_2)\) >                                                           \nn\\
                      &=&   \;\;\;  
                            A
                            \int d^8z
                          \d^8(z_1-z)  \nn\\ 
                      & &     \;\;\;\;\;\;\;(\dbs\ds 
                           + \frac{i}{2}\partial_{\alpha \dot{\alpha}}
                            \bar{D}^{\dot{\alpha}} D^{\alpha}  ) 
                             \d^8(z-z_2) \nn\\ 
                       &=&   \;\;\;  
                            A
                            \int d^8z
                          \d^8(z_1-z)  \nn\\                        
                       & &     \;\;\;\;\;\;\;\(\dbs\ds 
                           - \frac{1}{2}\Box  \) 
                             \d^8(z-z_2) \nn\\ \label{O12O12}                  \eeqs
\ni Eq (\ref{j12j12}) shows that all the component correlators are sum of $\d$-functions and space-time derivative acting on $\d$-functions, again consistent with the non-renormalization theorems. For the component scalars, it is easy to check the consistency of eq (\ref{O11O11b}) and eq (\ref{O12O12}) with the SU(4) $R$-symmetry. Eq (\ref{O11O11b}) and eq (\ref{O12O12}) give $<tr\p^1\p^1\,tr\pb_1\pb_1>=2<tr\p^1\pb_2\,\pb^1\p^2>=2A$. Now consider an infinitesimal SU(4) $R$-symmetry transformation of the component scalars $\p^i$ and $\pb_j$. In particular we consider a generator of SU(4) which acts on $\p$'s such that $\d\p^1=-\epsilon^{*}{\pb}_2, \;\d\pb_2=\epsilon\p^1$, and $\d\pb_1=-\epsilon\p^2$. Then SU(4) $R$-symmetry implies 
    
\beqs
0   & = &   \d<tr\p^1\pb_2\; tr\pb_1\pb_1>\nn\\
    & = &  -\epsilon^{*}<tr\pb_2\pb_2\;tr\pb_1\pb_1>+
             \epsilon <tr\p_1\p_1\;tr\pb_1\pb_1>
            -2\epsilon<tr\p_1\pb_2\;tr\pb_1\p_2>       
\eeqs 
\ni The first equality simply tells the vacum is $R$-symmetric. The first term of the second equality is zero giving the same relation as above.\\
\indent So far, we have considered two simple correlators, $<O^{11}O_{11}>$ and $<O^1_2O_1^2>$. One may consider different examples, but some of them can be obtained by inspection. $< (O_1^1-O_2^2)(O_1^1-O_2^2)>$ is such a case. The relevant graphs are the same as those of  $<O_1^2\;O^1_2>$ and are given by Fig.~\ref{fig:2all5}. The calculation will be almost identical giving the same result as in $<O^1_2 O_1^2>$ up to a factor of 2. The only subtlety is the sign of Fig.~\ref{fig:1212dif}. One might think that there is an additional minus sign for Fig.~\ref{fig:1212dif} of $< (O_1^1-O_2^2)(O_1^1- O_2^2)>$ coming from the expansion. However the color factor of Fig.~\ref{fig:2all5}b compensates for this minus sign.              

\begin{figure}[!ht]
\centerline{
\epsfxsize=4.5cm
\epsfbox{1212dift.epsi}
}
\caption{}
\label{fig:1212dif}
\end{figure}

Our final example is $<O^{11}O^{22}O^{33}>$. Performing $D$-algebra as indicated by the arrow, we have

\begin{figure}[!ht]
\centerline{
\epsfxsize=12cm
\epsfbox{pt3t1.epsi}
}
\caption{}
\label{fig:pt3t1}
\end{figure}

\ni which translates into

\be
S^{(2)}_{eff} =
           \int \prod_{i=1}^3d^4x_id^4\theta~
           \left[ \ds J_{11}~J_{22}~J_{33}~f_a  
       +
          J_{11}~ \ds J_{22}~ J_{33}~f_b  
       +
          D_{\alpha} J_{11}~ D^{\beta} J_{22}~ J_{33}~
            {{f_c}^\alpha}_\beta  \right]  \nn\\
\ee

\ni where $f_a,f_b,f_c$ are the corresponding Feynman diagrams on 
Fig.~\ref{fig:pt3t1}. To simplify further, we integrate over $\bar{\theta}^2$ and then partially integrate space time derivatives 

\beqs
S^{(2)}_{eff} =
         \int \prod_{i=1}^3d^4x_id^2\theta~
          \left[  \Box J_{11}~ J_{22}~ J_{33}~f_a 
       +
           J_{11}~ \Box J_{22}~ J_{33}~f_b 
       -
         \partial_{\alpha\dot{\rho}} J_{11}~
          \partial^{\beta\dot{\rho}} J_{22}~ J_{33}~
               {{f_c}^\alpha}_\beta   \right]\nn \\
\eeqs
\ni Integration by parts can be easily done diagrammatically:

\newpage

\begin{figure}[!ht]
\centerline{
\epsfxsize=12cm
\epsfbox{pt3t2.epsi}
}
\caption{}
\label{fig:pt3t2}
\end{figure}

\ni Thus we have the following result:

\beqs
S^{(2)}_{eff} 
         &=&
         \int \prod_{i=1}^3d^4x_id^2\theta~
       J_{11}(x_1,\theta) J_{22}(x_2,\theta) J_{33}(x_3,\theta)
       F(x_1,x_2,x_3) \nn
\eeqs       
                                                                                                                                                              \ni where 

\beqs
       F(x_1,x_2,x_3) 
       &=& 2
       \f{1}{(4\pi^2)^4} \f{1}{(x_1-x_2)^4}
        \f{1}{(x_2-x_3)^2}
        \f{1}{(x_1-x_3)^2}
         \nn \\
       &-&
      \f{1}{16\pi^{12}} \left[ \int d^4u
           \f{(u-x_2)^{\beta\dot{\rho}}}{(u-x_2)^4} 
           \f{(u-x_1)_{\alpha\dot{\rho}}}
              {(u-x_1)^4}\f{1}{(u-x_3)^2} \right]^2
\eeqs

\ni The second term in $F$ can be easily evaluated using method of \cite{EF} and it cancels the first term. Therefore $S_{eff}$=0. This shows at this order that all the component correlators including $<\p^1\p^1\,(2F^2\p^2+\psi^2\psi^2)\,(2F^3\p^3+\psi^3\psi^3)>$ which was considered in \cite{DFS} are zero verifying non-renormalization of three point functions.

\section{Gauge Choice Independence}

\indent Since the operators under consideration are gauge invariant, one expects the calculations to be gauge choice independent and as we see below, our results confirm the expectation. We consider Landau gauge. In addition to the Feynman gauge vector propagator, the \N=1 gauge propagator has the following extra piece,

\be
O\equiv\;(-1+\alpha)\frac{D^2\bar{D}^2+\bar{D}^2D^2}{ {\Box}^2 }
\label{Landau}
\ee

\ni In Landau gauge, all the graphs in Fig.~\ref{fig:2all1} need to be considered to compute $<O^{11}O_{11}>$ because the self-energy graphs are not zero. Since we have computed them in Feynman gauge, $\alpha=1$, what we must show is that all the extra contributions from eq(\ref{Landau}) with $\alpha=0$ cancel each other. Let's consider Fig~\ref{fig:2all1}a,b and c one by one. (Fig~\ref{fig:2all1}~d doesn't have a vector propagator. Therefore it need not be considered.) After rather straightforward $D$-algebra, Fig.~\ref{fig:2all1}a gives (From now on, a $\f{1}{\Box^2}$ acting on the vector lines is understood.)

\newpage
       
\begin{figure}[!ht]
\centerline{
\epsfxsize=8cm
\epsfbox{g1t.epsi}
}
\caption{}
\label{fig:gij}
\end{figure}

\ni Similarly for Fig.~\ref{fig:2all1}b,

\begin{figure}[!ht]
\centerline{
\epsfxsize=11cm
\epsfbox{g2t.epsi}
}
\caption{}
\label{fig:g2d-e}
\end{figure}

\ni The minus sign of the graph in the left hand side of the first equality  comes from the color factor. The second equality can be seen easily by writing out both sides explicitly.  Finally the Fig.~\ref{fig:2all1}c can be written

\begin{figure}[!ht]
\centerline{
\epsfxsize=8cm
\epsfbox{g3t.epsi}
}
\caption{}
\label{fig:g3}
\end{figure}

\ni Now it is obvious that the sum of the results in Fig.~\ref{fig:gij}, Fig.~\ref{fig:g2d-e}, and Fig.~\ref{fig:g3} adds up to zero verifying the gauge choice independence.    

It's more complicated to check the gauge independence of $<O^1_2O_1^2>$. In Landau gauge all the graphs in Fig.~\ref{fig:2all3} shown below contribute as well as the graphs in Fig.~\ref{fig:2all1}. However in this case again, all the extra contributions can be shown to cancel within themselves.

 \begin{figure}[!ht]
\centerline{
        \begin{minipage}[b]{2.5cm}
                \epsfxsize=2.5cm
                \epsfbox{2all4d.epsi}
        \end{minipage}
        \begin{minipage}[b]{2.5cm}
                \epsfxsize=2.5cm
                \epsfbox{2all4c.epsi}
        \end{minipage}                                                                 \begin{minipage}[b]{2.5cm}
                \epsfxsize=2.5cm
                \epsfbox{2all3a.epsi}
        \end{minipage} 
        \begin{minipage}[b]{2.5cm}
                \epsfxsize=2.5cm
                \epsfbox{2all4a.epsi}                     
            \end{minipage}    
}          
\caption{}       
\label{fig:2all3}
\end{figure}

\section{Conclusion}

In this article, we presented two loop computations of some gauge invariant two- and three- point correlators in \N=1 superspace. In accordance with the known non-renormalization theorems, they vanish for operators at separate points. For component scalars, we check the SU(4) $R$-symmetry. Since our results are consistent with \N=4 supersymmetry so far, we believe it is safe to use the Lagrangian eq (\ref{L}) for further computations.

\section*{Acknowledgements}        

We thank G. Chalmers and K. Schalm and especially M. Ro\v{c}ek for their enlightening discussions. F. Gonzalez-rey acknowledges support from NSF grant PHY-98-02551.  B. Kulik and I.Y. Park's work is partially supported by NSF grant PHY-97-22101.

\end{document}